\documentclass[12pt]{article}
\usepackage{amsmath, amsfonts, amsthm, amssymb}
\usepackage{geometry, graphicx, subfig, hyperref, cite, tikz, empheq}
\numberwithin{equation}{section}
\newcommand*\widefbox[1]{\fbox{\hspace{3.5em}#1\hspace{3.5em}}}

\begin{document}

\begin{titlepage}

\begin{flushright}
YITP-SB-17-34
\end{flushright}
\begin{center}
\vspace{1.0cm}
\Large{\textbf{Conformal manifolds: ODEs from OPEs}}\\
\vspace{0.8cm}
\small{\textbf{Connor Behan}}\\
\vspace{0.5cm}
\textit{C. N. Yang Institute for Theoretical Physics, Stony Brook University, \\ Stony Brook, NY 11794, USA}
\end{center}

\vspace{1.0cm}
\begin{abstract}
The existence of an exactly marginal deformation in a conformal field theory is very special, but it is not well understood how this is reflected in the allowed dimensions and OPE coefficients of local operators. To shed light on this question, we compute perturbative corrections to several observables in an abstract CFT, starting with the beta function. This yields a sum rule that the theory must obey in order to be part of a conformal manifold. The set of constraints relating CFT data at different values of the coupling can in principle be written as a dynamical system that allows one to flow arbitrarily far. We begin the analysis of it by finding a simple form for the differential equations when the spacetime and theory space are both one-dimensional. A useful feature we can immediately observe is that our system makes it very difficult for level crossing to occur.
\end{abstract}

\end{titlepage}

\tableofcontents

\section{Introduction}
In a $d$-dimensional conformal field theory, an exactly marginal operator is a primary scalar of dimension $d$ which does not pick up an anomalous dimension when it is added to the CFT as a deformation. The space of CFTs that can be reached in this way is referred to as a conformal manifold.\footnote{Continuous families of CFTs can arise in other ways as well. One example is the procedure in \cite{brrz17a, brrz17b} for constructing a line of nonlocal fixed points. Liouville theory may also be seen as a fixed line as one varies the central charge. These do not fit the definition of a conformal manifold because the lines are not traversed by deforming the CFT with a local operator.} When the points along it describe genuinely different theories, not related by a relabelling of operators, we call a conformal manifold non-trivial. All non-trivial conformal manifolds that have been discovered so far have some enhanced symmetry beyond the conformal group $SO(d + 1, 1)$. In particular, all known examples in $d \geq 3$ are supersymmetric. The reason for this could simply be better analytic control which makes it easier to discover new theories, or there could be a fundamental obstruction to non-supersymmetric conformal manifolds. It is therefore worthwhile to check if there are some universal features of the operator algebra that we can associate with the presence of exactly marginal operators. Just as the modern bootstrap \cite{rrtv08} seeks to determine whether a putative set of local operators can belong to a consistent conformal theory, there may be a test that can narrow down the space of CFTs to the space of conformal manifolds.

The original references in this subject proved non-renormalization theorems to discover conformal manifolds \cite{ggrs83, s88, ls95, s98, cy10}. To some extent, they did so by making explicit reference to a Lagrangian. Interestingly, some of these manifolds turn out to be strongly coupled at all points. There is also a growing body of work developing the non-perturbative understanding of these theories through the superconformal algebra \cite{gkstw10, ghksst16, gkosw17, bn15}. The short multiplets to which marginal operators must belong only exist in certain cases. Above 2D, these are $\mathcal{N} = 1, 2$ in 3D and $\mathcal{N} = 1, 2, 4$ in 4D \cite{cdi16}. In these algebras, additional requirements for finding a conformal manifold may be phrased in terms of representation theory and recombination rules. Various aspects of superconformal manifolds have been deduced from this line of reasoning including their dimensionality and the presence of complex structure. The orthogonal approach taken here will use conformal perturbation theory which does not rely on supersymmetry or the presence of a Lagrangian. The mnemonic
\begin{equation}
S \mapsto S + \int \textup{d}^dx g_i \hat{\mathcal{O}}^i \label{deformation}
\end{equation}
is merely an abstract statement for how we deform correlation functions. Considering a single direction on a conformal manifold that could be multi-dimensional, we will denote the associated marginal operator and its coupling by $\hat{\mathcal{O}}$ and $g$ respectively. An infinite family of constraints follows from setting $\beta(g)$, the running of the coupling, to zero. The two-loop term becomes a sum rule for even-spin CFT data analogous to the one in \cite{rrtv08}.

The local operators in a conformal manifold, even-spin or otherwise, obey many additional restrictions that require more work to state. Although there is no known way to tell if a set of scaling dimensions and OPE coefficients $\{ \Delta_i, \lambda_{ijk} \}$ is part of a conformal manifold, the framework of conformal perturbation theory holds promise in telling us whether two such sets can consistently be part of the \textit{same} conformal manifold. The key is that when there is a unique operator of each dimension, a set of differential equations exists for evolving $\{ \Delta_i(g), \lambda_{ijk}(g) \}$ from one value of $g$ to another. Subtleties arise when there is degeneracy and especially when there is more than one marginal operator. In this case, there is a non-trivial Zamolodchikov metric and the curvatures built up from it become interesting observables that affect how the equations for $\frac{\textup{d}\Delta_i}{\textup{d}g}$ and $\frac{\textup{d}\lambda_{ijk}}{\textup{d}g}$ must be defined \cite{rsz94, bnp17}. Even after we limit ourselves to a single marginal operator, these equations can only be written down once the appropriate conformal block expansions are known. This yields conformal block requirements that are much steeper than those in other CFT techniques. For comparison, we note that recent studies of the analytic bootstrap use conformal blocks with small external spin that only need to be evaluated in certain limits \cite{lmp16, hjk16, hlmpr16, lmp17}. Bounds from spinning correlators, recently found with the numerical bootstrap, use the full expressions, but again the external spin is at most 2 \cite{ikppsy16, ikpps17, dptv17, dkkps17}. The flow equations for conformal manifolds couple blocks of all internal and external Lorentz representations. For this reason they seem to be prohibitive in $d \geq 3$.

Nevertheless, we will see shortly that the system can in fact be analyzed sensibly in $d = 1$. The main result we have derived from this is that there is no level crossing for operators of the same symmetry. The absence of level crossing has long been predicted on general grounds but it remains a challenge to see how it is achieved. Because our argument assumes no degeneracy, we have not answered whether level crossing can occur when there is more than one marginal deformation. The program of applying our system to a known conformal manifold is similar in spirit to an algorithm that was recently developed for the numerical bootstrap. The authors of \cite{ps16} saturated a dimension bound and then obtained all other solutions to crossing on the edge of that bound through a set of evolution equations. We envision an algorithm that accomplishes the same thing except in a setting where the spectra belong to the same continuous line for a physical reason.

This paper is organized as follows. In section 2, we use the two-loop vanishing of $\beta(g)$ to derive the aforementioned sum rule that CFTs possessing a marginal operator must satisfy. We explain some technical features of it related to renormalization and the fact that spin-$\ell$ operators in $\hat{\mathcal{O}} \times \hat{\mathcal{O}}$ contribute with the sign $(-1)^{\ell / 2}$. This two-loop contribution to the beta function can be regarded as a one-loop shift in the OPE coefficient $\lambda_{\hat{\mathcal{O}}\hat{\mathcal{O}}\hat{\mathcal{O}}}$. In section 3, we generalize our construction to an arbitrary OPE coefficient $\lambda_{ijk}$. The infinitely many constraints that follow are recast into the explicit dynamical system in one dimension that we have advertised. The implications for nearly degenerate pairs of operators then follow from elementary scattering theory. Section 4 concludes and suggests future extensions of this work.

After these results were obtained, \cite{bbr17} appeared in which the same sum rule is derived. It should also be noted that the existence of a dynamical system for CFT data has been contemplated before this work. In particular, the ODEs we study were proposed in the talks \cite{h16, h17}.

\section{Two-loop constraints}
The standard framework for studying CFTs deformed by local operators is conformal perturbation theory \cite{z87, c96}. Defining $\mathcal{O}(\infty) = \lim_{x \rightarrow \infty} x^{2\Delta} \mathcal{O}(x)$, the order-$n$ correction to $\hat{\mathcal{O}}(\infty)$ is
\begin{equation}
\frac{g_0^n}{n!} \int \textup{d}^dx_1 \dots \textup{d}^dx_n \left < \hat{\mathcal{O}}(x_1) \dots \hat{\mathcal{O}}(x_n) \hat{\mathcal{O}}(\infty)\right > \; . \label{cpt-order-n}
\end{equation}
Integrals of this form generically have logarithmic divergences which should be regulated by a UV cutoff.\footnote{An approach using dimensional regularization instead was developed in \cite{bm14, bm16}.} Calling this cutoff $\Lambda$, it is enforced as a minimum distance between insertion points in the integration region: $|x_{ij}| < \Lambda^{-1}$. The divergences associated with shrinking this circle are removed from renormalized correlators by expressing them in terms of the coupling $g = \Lambda^{\hat{\Delta} - d} g_0$. The beta function is
\begin{equation}
\beta(g) = \frac{\textup{d}g}{\textup{d} \log\Lambda} = (\hat{\Delta} - d)g + \beta_2 g^2 + \beta_3 g^3 + \dots \; , \label{beta-function}
\end{equation}
which should vanish for marginal $\hat{\mathcal{O}}$. Putting our theory in a box of volume $V$ to handle an IR divergence, the one-loop and two-loop terms involving $\log \Lambda$ may be read off from
\begin{eqnarray}
\frac{1}{V} \int \textup{d}^dx_1 \textup{d}^dx_2 \left < \hat{\mathcal{O}}(x_1) \hat{\mathcal{O}}(x_2) \hat{\mathcal{O}}(\infty) \right > &\sim& -2 \beta_2 \log \Lambda \nonumber \\
\frac{1}{V} \int \textup{d}^dx_1 \textup{d}^dx_2 \textup{d}^dx_3 \left < \hat{\mathcal{O}}(x_1) \hat{\mathcal{O}}(x_2) \hat{\mathcal{O}}(x_3) \hat{\mathcal{O}}(\infty) \right > &\sim& -6 \beta_3 \log \Lambda + 6 \beta_2^2 \log^2 \Lambda \; . \label{quantum-integrals}
\end{eqnarray}
In the first integral, the OPE with $x_2 \rightarrow x_1$ tells us that
\begin{equation}
\beta_2 = -\frac{S_{d - 1}}{2} \lambda_{\hat{\mathcal{O}}\hat{\mathcal{O}}\hat{\mathcal{O}}} \; . \label{beta2}
\end{equation}
The second integral is more interesting as it involves a four-point function. Recently, there has been interest in approximating it using data from the numerical bootstrap \cite{ks16, brrz17a, brrz17b}. A logarithmic divergence arises by letting $x_2$ and $x_3$ approach $x_1$ while remaining of the same order. Performing a conformal transformation, we may write $\beta_3$ as a single integral of $\left < \hat{\mathcal{O}}(0) \hat{\mathcal{O}}(x) \hat{\mathcal{O}}(\hat{e}) \hat{\mathcal{O}}(\infty) \right >$ where $\hat{e}$ is an arbitrary unit vector. For each relevant operator in the OPE $\hat{\mathcal{O}} \times \hat{\mathcal{O}}$, there is a power-law singularity. Subtracting these,
\begin{equation}
\beta_3 = -\frac{S_{d-1}}{6} \int \textup{d}^dx \left [ \left < \hat{\mathcal{O}}(0) \hat{\mathcal{O}}(x) \hat{\mathcal{O}}(\hat{e}) \hat{\mathcal{O}}(\infty) \right > - \sum_{\Delta < d} \lambda^2_{\hat{\mathcal{O}}\hat{\mathcal{O}}\mathcal{O}} \left ( \frac{1}{|x|^\Delta} + \frac{1}{|x|^{2d - \Delta}} + \frac{1}{|\hat{e} - x|^{2d - \Delta}} \right ) \right ] \label{beta3a}
\end{equation}
is the net result.\footnote{Equivalently, one could omit $\Delta = 0$ from the sum and then invoke some notation to change the four-point function to the \textit{connected} four-point function.} 
\begin{figure}[t!]
\centering
\includegraphics[scale=0.6]{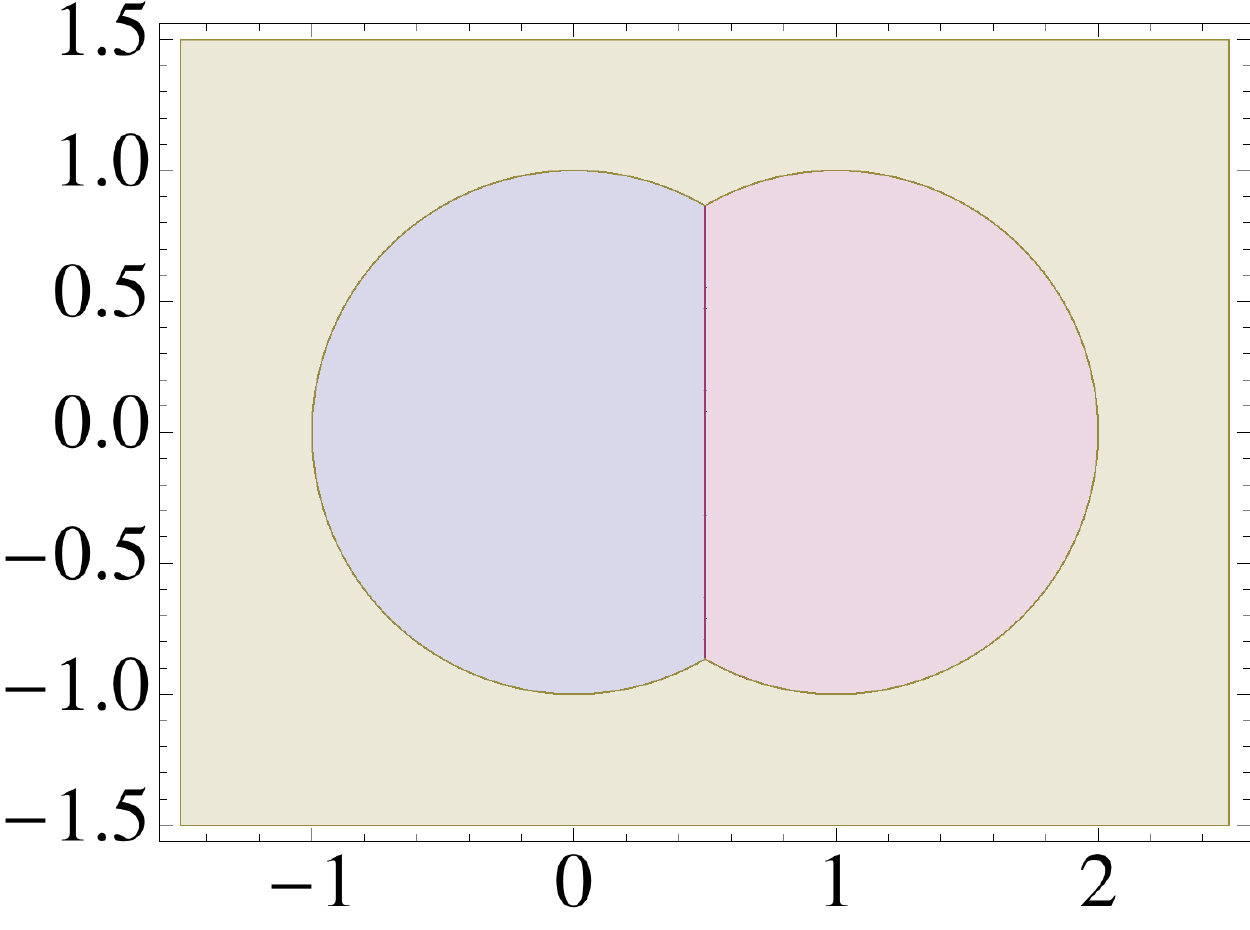}
\caption{Once we send our four points to $(0, z, 1, \infty)$, $\mathcal{R}_{12}$, $\mathcal{R}_{23}$ and $\mathcal{R}_{13}$ map to the blue, red and yellow $z$-plane regions respectively. We have used a change of variables to give all integrals the blue domain, which we denote by $\mathcal{R}$.}
\label{stu-regions}
\end{figure}
There is a different form of (\ref{beta3a}) that will be more useful for our purposes \cite{ks16}. It comes from writing (\ref{quantum-integrals}) as an integral over $\mathcal{R}_{12} \cup \mathcal{R}_{23} \cup \mathcal{R}_{13}$ where $\mathcal{R}_{ij}$ means that $|x_{ij}|$ is smaller than the other two distances. These are precisely the regions of optimal convergence for conformal block expansions in the $s$, $t$ and $u$ channels. Figure \ref{stu-regions} plots them for our desired kinematics. By covariance of the correlator, they can all be swapped for the region $\mathcal{R} \equiv \{ |x| < 1, |\hat{e} - x| \}$ where the only potential singularity is at the origin. With this in mind,
\begin{equation}
\beta_3 = -3 \frac{S_{d-1}}{6} \sum_{\mathcal{O}} \lambda^2_{\hat{\mathcal{O}}\hat{\mathcal{O}}\mathcal{O}} \biggl. \int_{\mathcal{R}} \textup{d}^dx |x|^{-2d} G_{\mathcal{O}}(u, v) \biggl |_{\mathrm{reg}} \; . \label{beta3b}
\end{equation}
The factor of 3 reflects the crossing symmetry of the four-point function. Later, when we deal with mixed correlators, we will have to treat each permutation separately. By setting the terms above to zero,
\begin{empheq}[box=\fbox]{align}
& \hat{\Delta} = d \nonumber \\
& \lambda_{\hat{\mathcal{O}}\hat{\mathcal{O}}\hat{\mathcal{O}}} = 0 \nonumber \\
& \sum_{\mathcal{O}} \lambda^2_{\hat{\mathcal{O}}\hat{\mathcal{O}}\mathcal{O}} \biggl. \int_{\mathcal{R}} \textup{d}^dx |x|^{-2d} G_{\mathcal{O}}(u, v) \biggl |_{\mathrm{reg}} = 0 \label{sum-rule}
\end{empheq}
are the conditions that a conformal manifold imposes on the operator algebra. For the rest of this section, we will focus on the non-trivial line of (\ref{sum-rule}) and refer to it as the \textit{sum rule}.
\begin{figure}[t!]
\subfloat[][Scalars, 2D]{\includegraphics[scale=0.4]{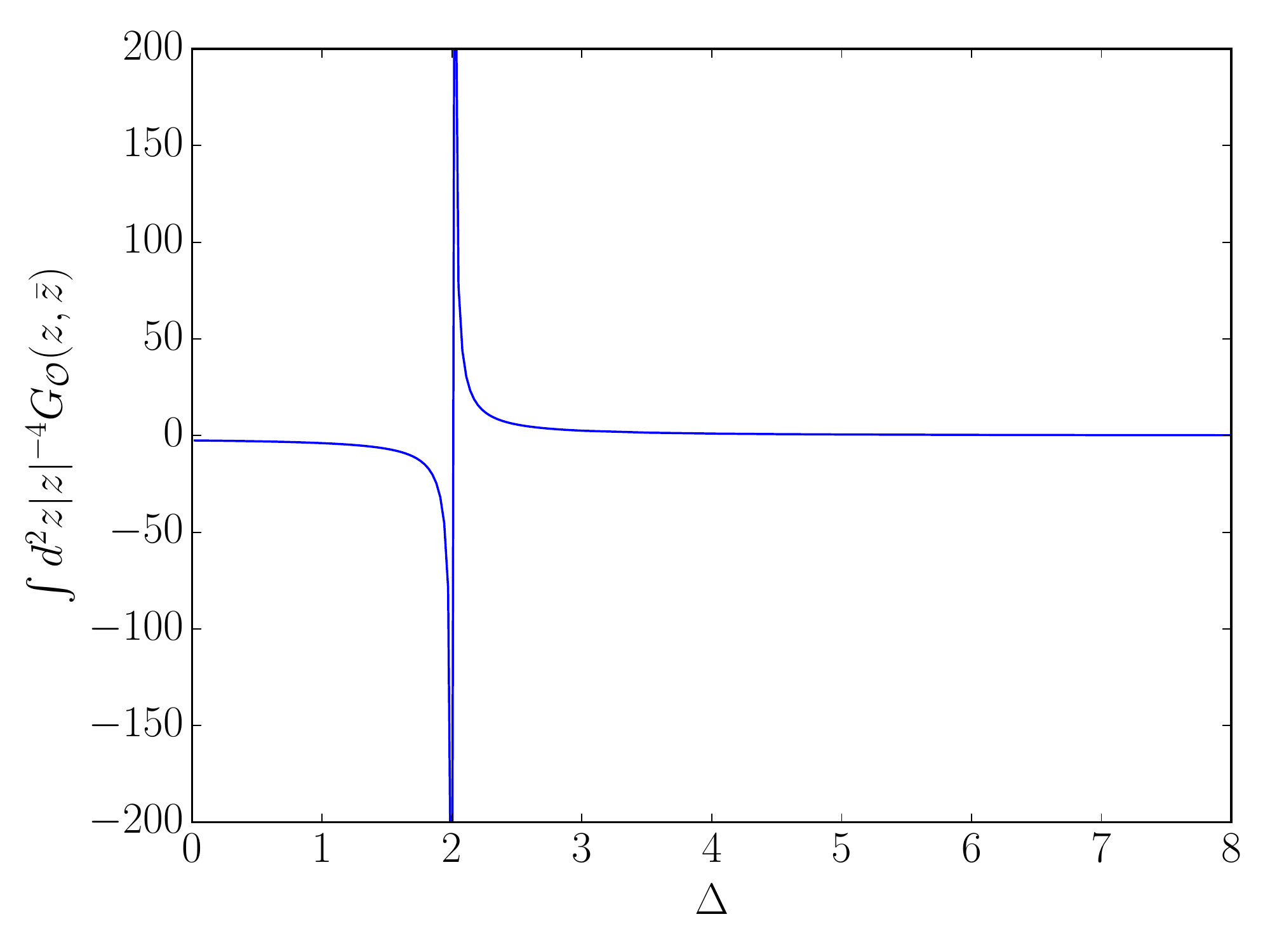}}
\subfloat[][Tensors, 2D]{\includegraphics[scale=0.4]{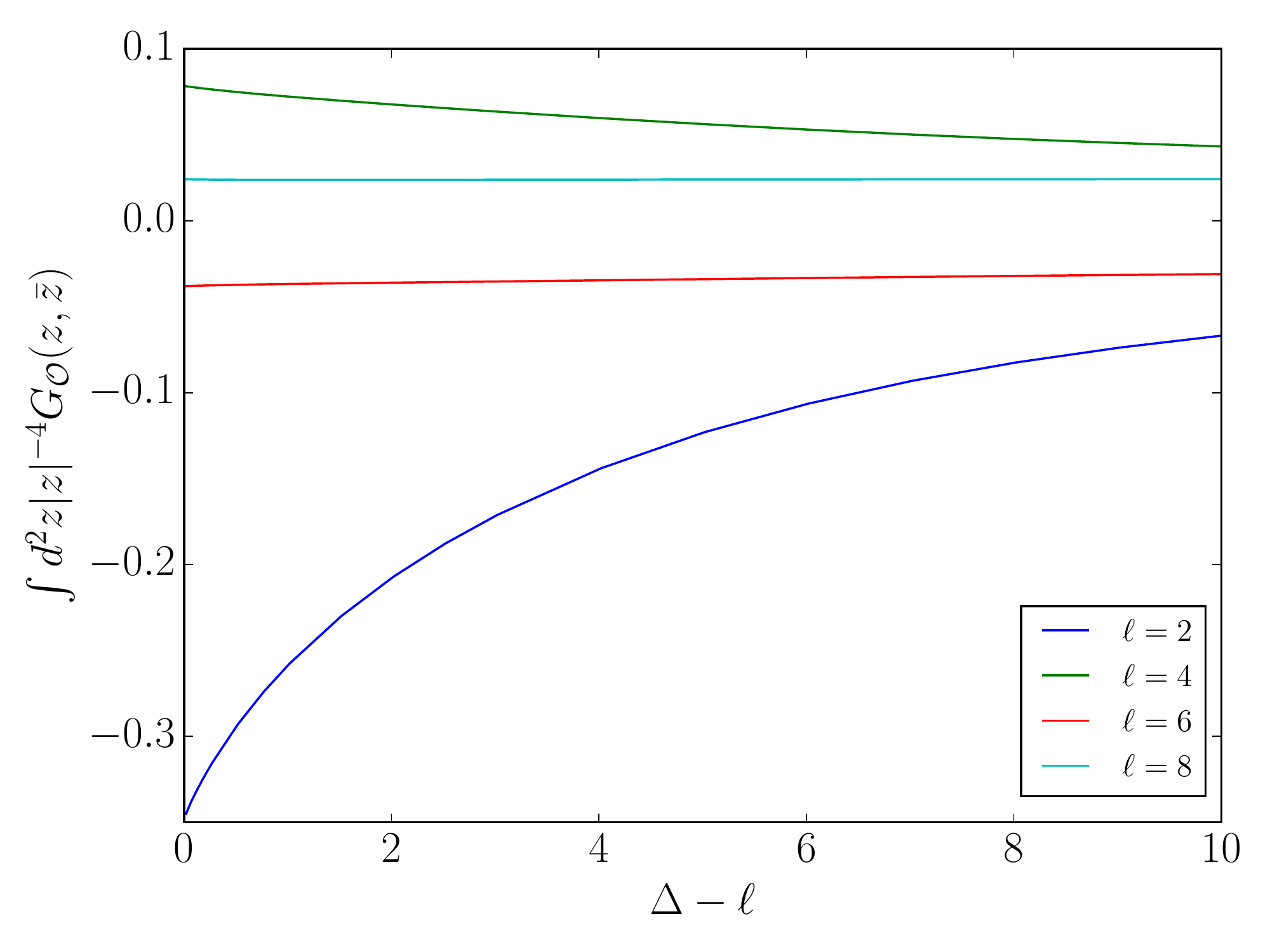}}
\caption{Plots showing how a primary operator in $\hat{\mathcal{O}} \times \hat{\mathcal{O}}$ contributes to the beta function. These follow from a numerical integral but they may be obtained analytically in one dimension.}
\label{plot-tensor}
\end{figure}
As a first step in understanding the ingredients of the sum rule, we may plot the conformal block integrals as functions of the exchanged dimension $\Delta$. Figure \ref{plot-tensor} does this for $d = 2$ but the same basic features appear in all dimensions. Looking at the scalars, we see that relevant and irrelevant operators contribute with opposite signs. The discontinuity at $\Delta = d$ arises because the counterterm $\frac{1}{|x|^\Delta}$, introduced to cure a UV divergence, becomes IR divergent as well at marginality. Higher spin operators, which exist for $d \geq 2$, appear to have the same sign for all $\Delta$. However, they alternate with spin according to $\ell \; (\mathrm{mod} \; 4)$. This phenomenon was noted in \cite{bbr17} which also included plots for $d = 4$.

\subsection{Ambiguities in the sum rule}
\begin{figure}[h!]
\centering
\subfloat[][]{\includegraphics[scale=0.4]{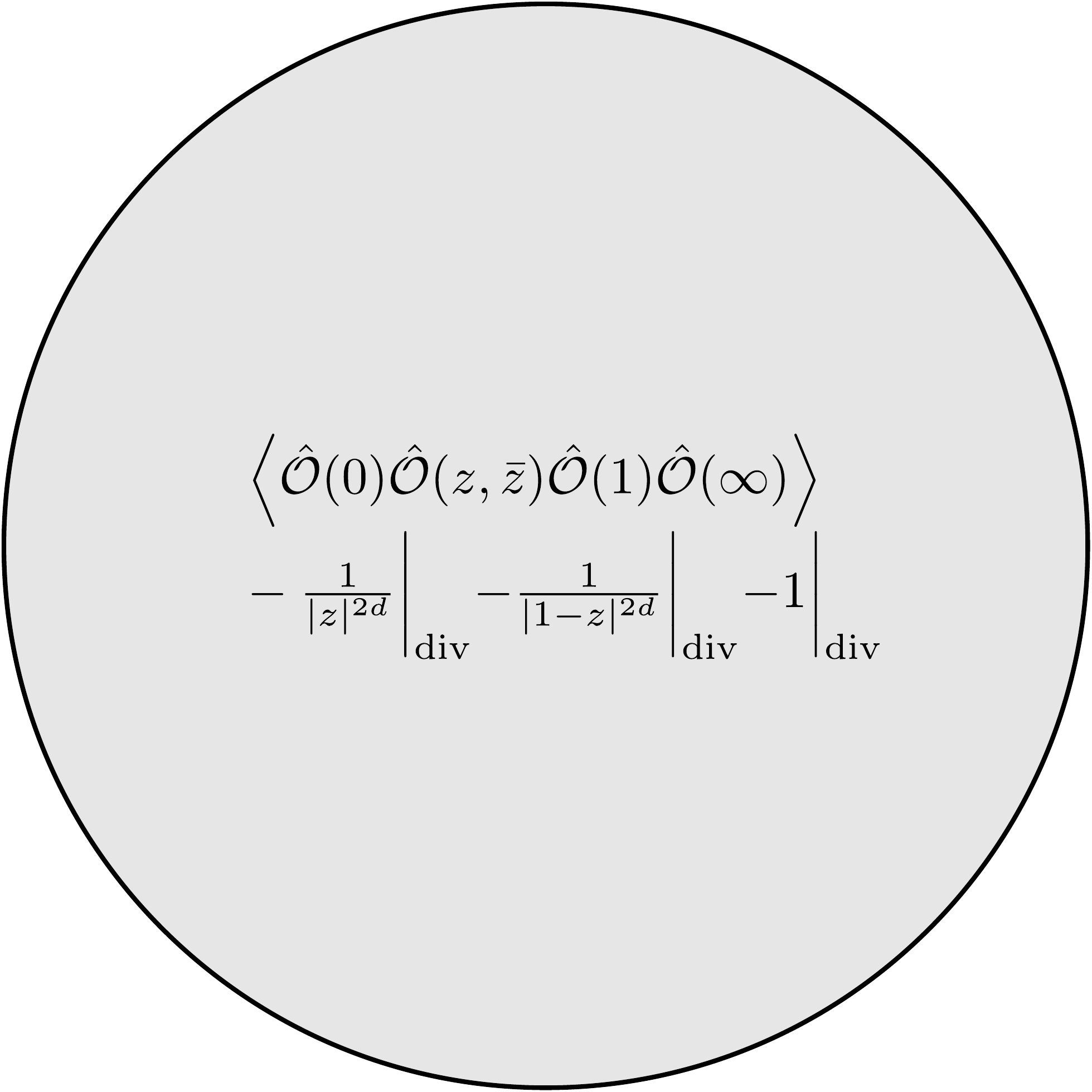}}
\subfloat[][]{\includegraphics[scale=0.4]{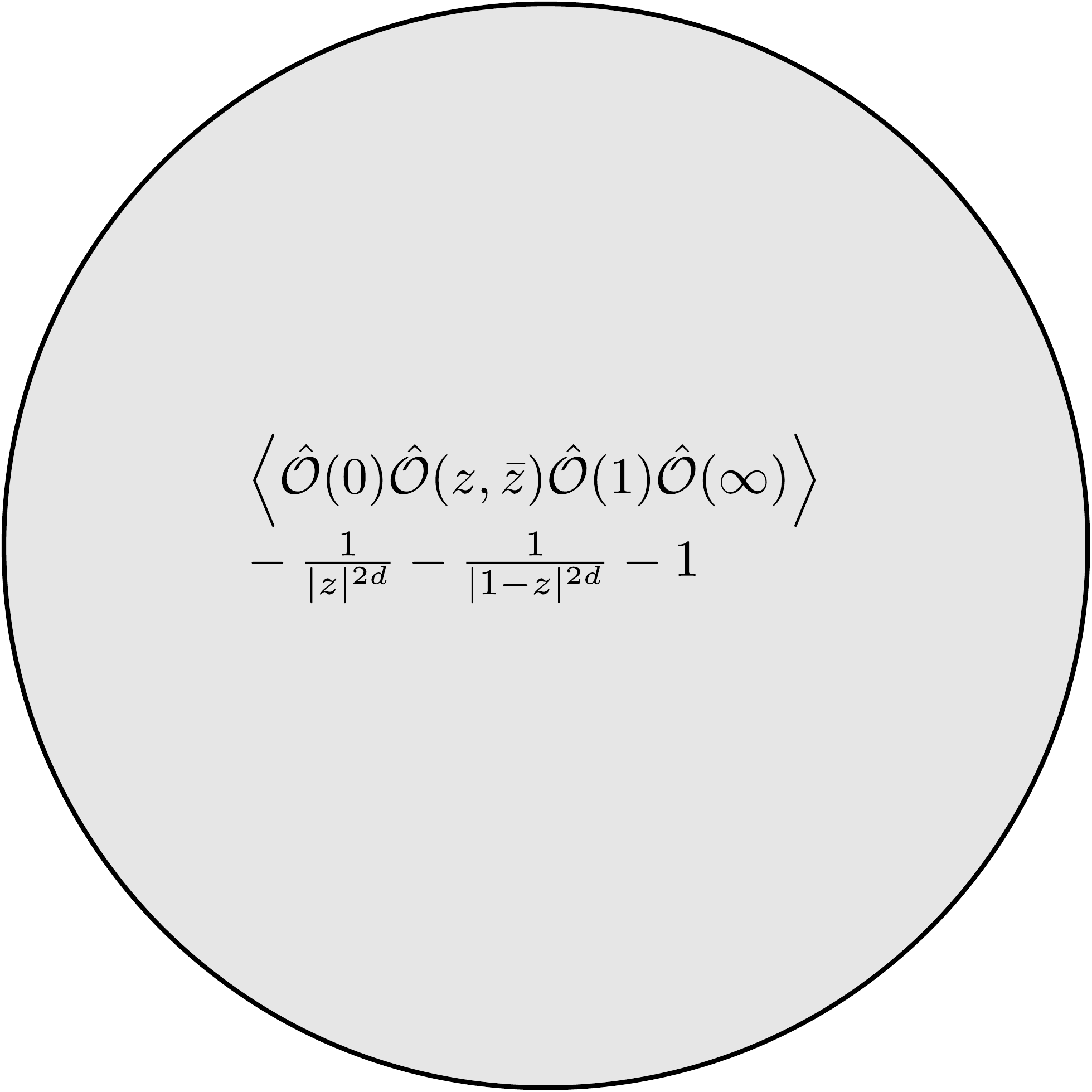}}
\caption{We represent $\mathbb{R}^d$ as a blob with the function to be integrated inside it. The left and right choices both compute the two-loop beta function. Because $\mathbb{R}^d$ has no boundary, the integrated power-laws being subtracted are equal to their divergent parts.}
\label{circle1}
\end{figure}

\begin{figure}[h!]
\centering
\subfloat[][Equivalent to Figure \ref{circle1} (a)]{\includegraphics[scale=0.4]{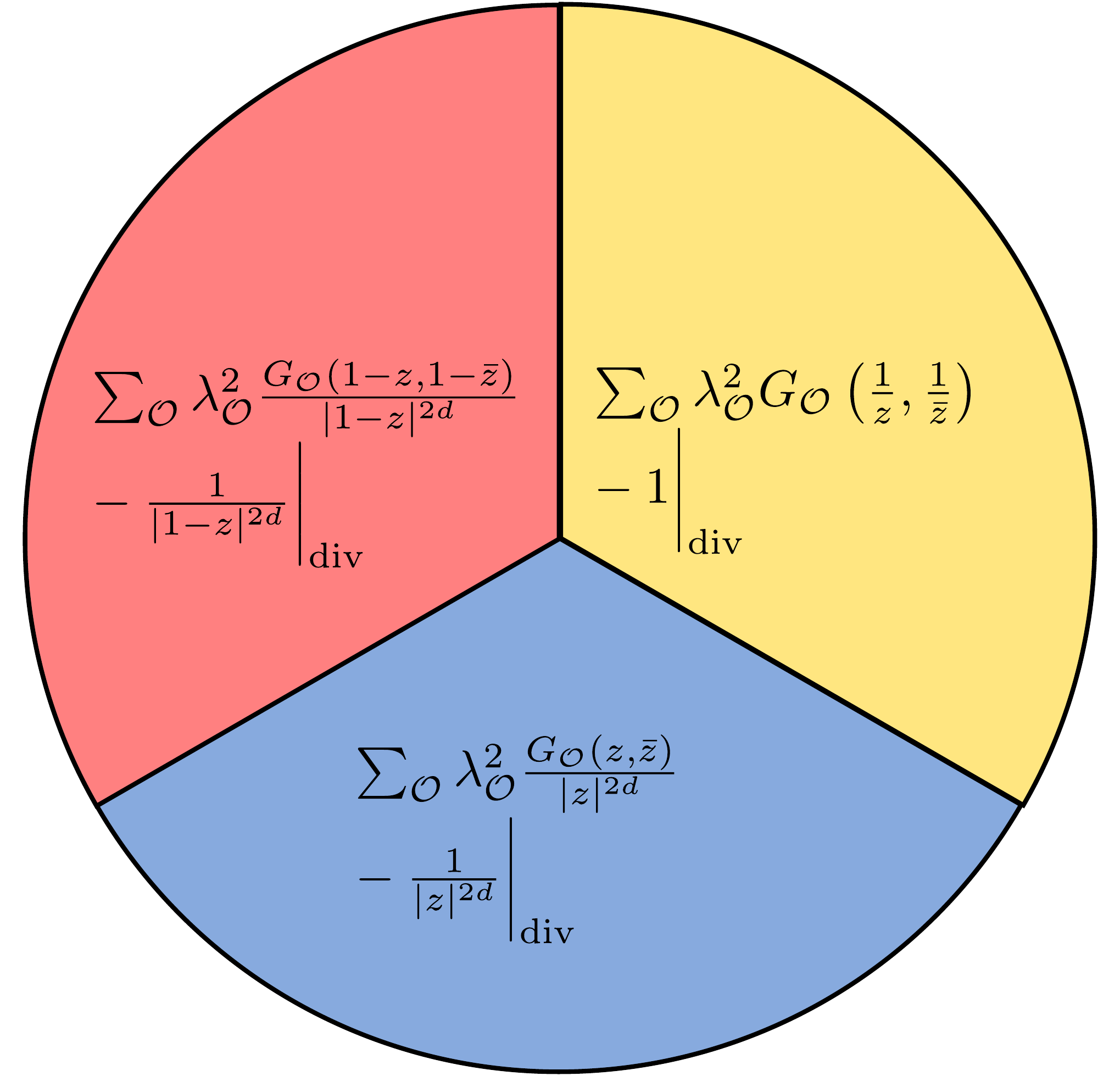}}
\subfloat[][Equivalent to Figure \ref{circle1} (b)]{\includegraphics[scale=0.4]{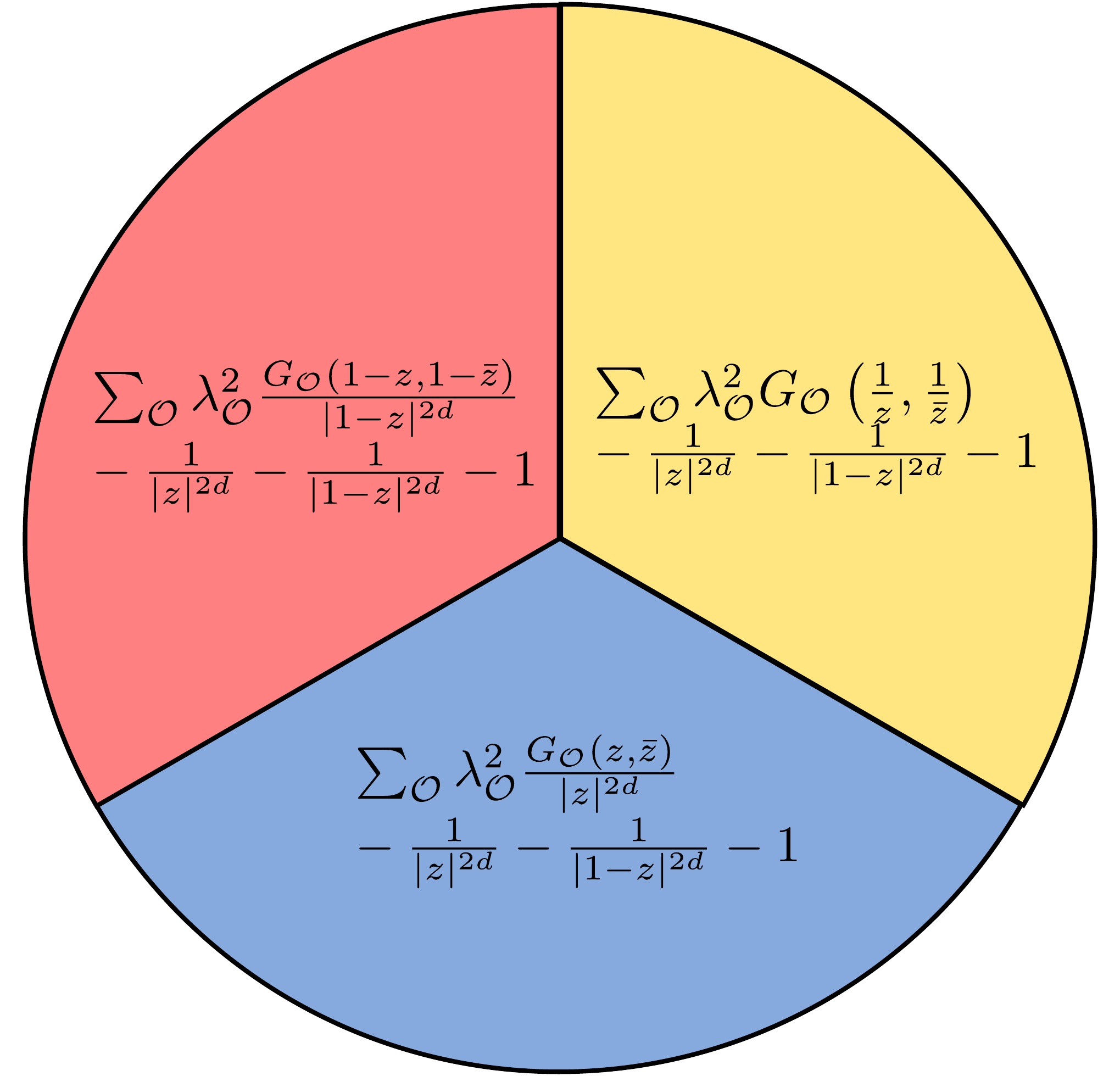}}
\caption{The cartoons obtained by splitting Figure \ref{circle1} into $s$, $t$ and $u$ channel regions. In both cases, the identity block is not annihilated. A divergence subtraction is now no longer the same as a full subtraction. In particular, removing $\mathrm{div}$ everywhere in the left blob would not compute a physical quantity.}
\label{circle2}
\end{figure}

An important feature of the scalar plot in Figure \ref{plot-tensor} is that it does not pass through the origin. This means that in the sum rule derived from a connected four-point function in (\ref{beta3a}), the identity term is finite. Although this might seem strange at first sight, it is an inevitable consequence of using counterterms in bounded regions. There are different ways to do this when crossing symmetry is not manifestly satisfied, leading to a freedom in how the sum rule is presented. The point is that the minimal subtraction we have adopted for an integrated conformal block is not the same as subtracting terms of the form $\frac{1}{|x|^\Delta} + \frac{1}{|x|^{2d - \Delta}} + \frac{1}{|\hat{e} - x|^{2d - \Delta}}$ given in (\ref{beta3a}). If we truly wanted to do the latter, our integrals of conformal blocks would need to be regulated by subtracting finite terms as well. In this discussion, we will explore the differences between these two choices. It should also be clear that the number of choices is infinite since we are free to apply any linear combination of the two above. We should emphasize that this is not a \textit{physical} ambiguity. Data solving one version of the sum rule will automatically solve another if it comes from a crossing symmetric theory. This reflects the fact that the beta function is well defined within our scheme. It is even scheme independent at two loops, as explained in \cite{ks16}. Nevertheless, it represents an ambiguity in characterizing the beta function term of an \textit{abstract} conformal multiplet.

For simplicity, suppose that the only relevant scalar in $\hat{\mathcal{O}} \times \hat{\mathcal{O}}$ is the identity. In this case, the right hand side of Figure \ref{circle1} is a cartoon that represents the integration done in (\ref{beta3a}). To change prescriptions, it will be helpful to distinguish between a subtracted function and its isolated divergence. We may consider a subtraction of $\frac{1}{|z|^{2d}}$ from $f(z)$, integrated over an $r$-ball for instance.
\begin{eqnarray}
\int_{B(r)} \textup{d}^dz f(z) - \frac{1}{|z|^{2d}} &=& \int_{B(r)} \textup{d}^dz f(z) - \frac{S_{d-1}}{d} (r^{-d} - \Lambda^d) \nonumber \\
\int_{B(r)} \textup{d}^dz f(z) - \biggl. \frac{1}{|z|^{2d}} \biggl |_{\mathrm{div}} &=& \int_{B(r)} \textup{d}^dz f(z) + \frac{S_{d-1}}{d} \Lambda^d \label{power-vs-divergence}
\end{eqnarray}
The expression that subtracts the full power-law includes an extra $\frac{S_{d-1}}{d} r^{-d}$ compared to the expression that only subtracts the divergence. Clearly, $r \rightarrow \infty$ makes these procedures equivalent, explaining why both halves of Figure \ref{circle1} are the same.

As soon as we expand in conformal blocks, we must partition space into the blue, red and yellow regions from Figure \ref{stu-regions}. These are represented in Figure \ref{circle2} by the same colors. The type of subtraction being performed on the left hand side is the choice made in this work. It makes use of the fact that divergences are localized around special points, allowing us to keep only one in each region. It is now clear that power-laws present in the blocks for relevant scalars will not be fully removed. After divergences are subtracted, finite boundary terms will remain. Another choice that we could have made is the approach of the right hand side --- subtracting the same original counterterm in all three channels. The equivalence between the two choices relies on crossing symmetry. In this case, a power-law like $\frac{1}{|z|^{2d}}$ is fully cancelled but the additional appearances of $1$ and $\frac{1}{|1 - z|^{2d}}$ mean that we are subtracting too much. Finite parts will thus persist unless we find a natural way to pair these counterterms with the blocks of irrelevant operators instead. In this regard, the lightcone bootstrap has successfully matched crossed-channel singularities to infinite towers of operators in the direct channel \cite{fkps12, kz12}. However, these are the well known towers of double-twist operators which only exist asymptotically.

The observation that the two sides of Figure \ref{circle2} are equivalent is not new. The divergence subtraction on the left hand side was referred to as ``method 2'' in \cite{brrz17a}. ``Method 1'' was discussed in a context where the four-point function was known exactly, but it can also be applied in cases where we have to use the conformal block expansion. This would make it identical to the right hand side of Figure \ref{circle2}.

\subsection{The alternating sign}
We now turn to the question of why the contributions plotted in Figure \ref{plot-tensor} have signs that alternate with spin. Although a general proof eludes us, we may show that the correct sign is predicted by the large-$\Delta$ limit. It is convenient to express everything in terms of the radial co-ordinate
\begin{eqnarray}
\rho &=& \frac{z}{(1 + \sqrt{1 - z})^2} \nonumber \\
r &=& |\rho| \nonumber \\
\eta &=& \cos \arg \rho \; . \label{radial-coord}
\end{eqnarray}
For the large-$\Delta$ block, we use
\begin{eqnarray}
G_{\Delta, \ell}(r, \eta) &=& \frac{\ell!}{(2\nu)_\ell} \frac{(4r)^\Delta C^\nu_\ell(\eta)}{(1 - r^2)^\nu \sqrt{(1 + r^2)^2 - 4r^2\eta^2}} \left ( 1 + O \left ( \frac{1}{\Delta} \right ) \right ) \nonumber \\
\nu &=& \frac{d - 2}{2} \label{large-delta-block}
\end{eqnarray}
which is the entire part from the meromorphic expansion in \cite{kps13}. It is easy to check that the region $\mathbb{C} \setminus (1, \infty)$ for $z$ maps to the unit circle for $\rho$ \cite{hr13}. Therefore $G_{\Delta, \ell}(r, \eta)$ vanishes for $\Delta \rightarrow \infty$ and so must an integral of it over a bounded region. Another thing to check is that $\mathcal{R}$, the region that looks like a cutoff circle, maps to $\{ (r, \eta) | -1 < \eta < 1, 0 < r < r_*(|\eta|) \}$ where $r_*(|\eta|)$ is the smaller solution of $r^2 - 4r + 1 = 2r|\eta|$ \cite{ks16}. With a little bit of work to find the measure, our integral is
\begin{equation}
I = \frac{\ell!}{(2\nu)_\ell} \int_{-1}^1 \textup{d}\eta \int_0^{r_*(|\eta|)} \textup{d}r \sqrt{(1 + r^2)^2 - 4r^2\eta^2} (4r)^{\Delta - d - 1} (1 - r^2)^\nu (1 - \eta^2)^{\nu - \frac{1}{2}} C^\nu_\ell(\eta) \; . \label{large-delta-integral}
\end{equation}
Everything in the integrand except the Gegenbauer polynomial is sign-definite and peaks at $\eta = 0$. It therefore seems that the sign of $I$ should be controlled by the sign of $C^\nu_\ell(0)$. This is indeed $(-1)^{\ell / 2}$ as can be seen in various ways. Perhaps most easily, we can just use the fact that each Gegenbauer is an even polynomial with $\ell$ zeros on the unit interval. The fact that $C^\nu_\ell(\pm 1) > 0$ then gives us the right sign for $C^\nu_\ell(0)$. To see why this sign persists, consider
\begin{equation}
\int_{-1}^1 \textup{d}\eta \frac{C^\nu_\ell(\eta)}{C^\nu_\ell(0)} f(|\eta|) (1 - \eta^2)^{\nu - \frac{1}{2}} \; . \label{integral-step1}
\end{equation}
If $f$ is identically $1$, then it is a special case of a Gegenbauer polynomial and (\ref{integral-step1}) vanishes by orthogonality. This means that the positive sign near $\eta = 0$ is exactly cancelling the negative signs that appear elsewhere. The integral should then become more positive once we change $f$ to a positive function that peaks at zero and decays in either direction. We may check that this holds for
\begin{equation}
f(|\eta|) = \int_0^{r_*(|\eta|)} \textup{d}r \sqrt{(1 + r^2)^2 - 4r^2\eta^2} (4r)^{\Delta - d - 1} (1 - r^2)^\nu \; , \label{decaying-profile}
\end{equation}
but this only proves our suspicion when $\ell \leq 2$. For higher spins, we will need information about $f$ beyond monotonicity. A saddle point evaluation of (\ref{decaying-profile}) leads to
\begin{equation}
f(|\eta|) \approx \frac{4^{\Delta - d}}{\Delta - d - 1} \sqrt{1 + |\eta|} (1 - r_*(|\eta|)^2)^\nu r_*(|\eta|)^{\Delta - d + 1} \; . \label{decaying-saddle}
\end{equation}
For sufficiently large $\Delta$, the increasing function $\sqrt{1 + |\eta|} (1 - r_*(|\eta|)^2)^\nu$ does not spoil the decrease of $r_*(|\eta|)^{\Delta - d + 1}$. Therefore, we may consider the parts of (\ref{integral-step1}) on either side of the first zero of $C^\nu_\ell(\eta)$. Call this point $\eta_0$. For the $\eta > \eta_0$ part of (\ref{integral-step1}) which includes negative contributions, the crudest underestimate is
\begin{eqnarray}
I_> &\geq& - \frac{4^{\Delta - d}}{\Delta - d - 1} K_> r_*(\eta_0)^{\Delta - d + 1} \nonumber \\
K_> &=& \left | \mathrm{min} C^\nu_\ell(\eta) / C^\nu_\ell(0) \right | (1 - \eta_0) \sqrt{1 + \eta_0} (1 - r_*(\eta_0)^2)^\nu \; . \label{integral-step2}
\end{eqnarray}
We may underestimate the positive contribution from $\eta \in [0, \eta_0]$ by just taking $\eta \in \left [ 0, \frac{\eta_0}{2} \right ]$. This yields
\begin{eqnarray}
I_< &\geq& \frac{4^{\Delta - d}}{\Delta - d - 1} K_< r_* \left ( \frac{\eta_0}{2} \right )^{\Delta - d + 1} \nonumber \\
K_< &=& \left | C^\nu_\ell \left ( \frac{\eta_0}{2} \right ) / C^\nu_\ell(0) \right | \frac{\eta_0}{2} \sqrt{1 + \frac{\eta_0}{2}} \left ( 1 - r_* \left ( \frac{\eta_0}{2} \right )^2 \right )^\nu \; . \label{integral-step3}
\end{eqnarray}
Since $K_>$ and $K_<$ are independent of $\Delta$, it is easy to choose $\Delta$ such that $[r_* \left ( \frac{\eta_0}{2} \right ) / r_*(\eta_0) ]^{\Delta - d + 1}$ is larger than $K_> / K_<$.

\subsection{Realizations}
The simplest conformal manifold is probably the compactified free boson in two dimensions.\footnote{We use the term ``manifold'' loosely, as the moduli space is not smooth at the self-dual radius.} While the absence of a beta function in a free theory hardly needs to be stated, it is instructive to discuss this model from symmetry considerations alone. At a generic radius, the symmetry is given by two copies of the affine $U(1)$ algebra. The current $J$ and its modes $a_n$ satisfy
\begin{eqnarray}
J(z) J(w) &=& \frac{1}{(z - w)^2} + \dots \nonumber \\
\left [ a_n, a_m \right ] &=& n \delta_{m + n, 0} \; . \label{u1}
\end{eqnarray}
The object $T = :J^2:$, known as the Sugawara stress tensor, has all of its properties determined from (\ref{u1}). Specifically,
\begin{eqnarray}
T(z) T(w) &=& \frac{1}{2(z - w)^4} + \frac{2T(w)}{(z - w)^2} + \frac{\partial T(w)}{z - w} + \dots \nonumber \\
L_m &=& \frac{1}{2} \sum_{n \in \mathbb{Z}} : a_{m - n} a_n : \label{virasoro}
\end{eqnarray}
which implies the Virasoro algebra for central charge $1$. This leads to a simple relation between the charges and conformal weights of primary operators; $h = q^2$. We therefore see that $J \bar{J}$ is a marginal deformation because the two pieces have charge $\pm 1$. 

Cardy found this model through a bottom-up approach while searching for a fixed line that did not require finely tuned OPE coefficients. In \cite{c87}, he showed that
\begin{eqnarray}
\left < \hat{\mathcal{O}}(0)\hat{\mathcal{O}}(z, \bar{z})\hat{\mathcal{O}}(1)\hat{\mathcal{O}}(\infty) \right > &=& |z|^{-4} G(z, \bar{z}) \label{cardy-4pt} \\
&=& 2\Re \left [ \frac{1}{z^2} + \frac{1}{(1 - z)^2} + \frac{1}{\bar{z}^2 (1 - z)^2} \right ] + \frac{1}{|z|^4} + \frac{1}{|1 - z|^4} + 1 \nonumber
\end{eqnarray}
is the unique crossing-symmetric four-point function with a vanishing regulated integral whose singularities involve only the Virasoro identity block. We have separated the connected and disconnected pieces above. The holomorphic factorization looks like
\begin{eqnarray}
G(z, \bar{z}) &=& g(z) g(\bar{z}) \nonumber \\
g(z) &=& \left ( \frac{z^2 - z + 1}{1 - z} \right )^2 = 1 + 2z^2 + \sum_{k = 2}^\infty k z^{k + 1} \; . \label{cardy-holomorphic}
\end{eqnarray}
The expansion of (\ref{cardy-holomorphic}) into $SL(2; \mathbb{R})$ blocks is
\begin{eqnarray}
g(z) &=& 1 + \sum_{n = 1}^\infty c_n z^{2n} {}_2F_1(2n, 2n; 4n; z) \nonumber \\
c_n &=& \frac{2n - 1}{4^{n - 1}} \frac{(2n)!}{(4n - 3)!!} \; , \label{cardy-blocks}
\end{eqnarray}
yielding explicit OPE coefficients in $\hat{\mathcal{O}} \times \hat{\mathcal{O}}$.\footnote{We have guessed (\ref{cardy-blocks}) with the help of OEIS \cite{oeis}. However, a proof should be possible with the technology of \cite{hv17}.} A spinning operator (labelled by non-negative integers $n, \bar{n}$) has $\lambda^2_{\hat{\mathcal{O}}\hat{\mathcal{O}}\mathcal{O}} = 2c_nc_{\bar{n}}$, while a scalar operator has $\lambda^2_{\hat{\mathcal{O}}\hat{\mathcal{O}}\mathcal{O}} = c_n^2$. Using these OPE coefficients, one can proceed to integrate the $SL(2; \mathbb{C})$ blocks in (\ref{cardy-4pt}). It should be no surprise that this verifies the sum rule (\ref{sum-rule}). The only scalar of $\Delta < 2$ in this Virasoro identity block is the quasiprimary $I$ itself. This would make it a \textit{dead-end CFT} if we took the unusual step of regarding this subsector of the free boson as a theory in its own right. Because no relevant deformations contribute to the beta function, the model owes marginality to the appearance of both signs in Figure \ref{plot-tensor}.

The known examples of conformal manifolds above $d = 2$ are superconformal manifolds. Let us discuss 4D theories in decreasing order of supersymmetry. The $\mathcal{N} = 4$ theories will have an exactly marginal operator whenever they have a stress-energy tensor. Both of these are in a multiplet whose primary transforms in the $\textbf{20}^\prime$ representation of the $SU(4)$ R-symmetry. Reassuringly, there is no way for this to recombine with another multiplet. There is a well known conjecture that all $\mathcal{N} = 4$ SCFTs fall into the Super Yang-Mills class. The situation in $\mathcal{N} = 2$ is similar with classical marginality implying exact marginality again due to recombination rules. The difference is that local $\mathcal{N} = 2$ SCFTs do not automatically require the multiplet for marginal operators to be present. Indeed, known $\mathcal{N} = 2$ SCFTs include conformal manifolds but also a large zoo of isolated fixed points \cite{t13}. Finally, classically marginal operators are ubiquitous in $\mathcal{N} = 1$ theories as descendants of scalar chiral primaries. These primaries obey $\Delta = \frac{3}{2}|r|$, where we are interested in $r$-charge 2 to get the multiplet of a superpotential. To prevent this from recombining with a conserved current multiplet, one needs extra input such as the mechanism described in \cite{gkstw10}.

Since we must have a solved theory to fully apply (\ref{sum-rule}), we are essentially limited to examples like free $\mathcal{N} = 4$ SYM or free $\mathcal{N} = 2$ SQCD with the right matter content to make it conformal. What might be more interesting is checking the contribution of particular supermultiplets. BPS multiplets, which have fixed dimension whenever they appear, could be treated once. When an unprotected multiplet has correlators of its descendants determined by those of the primary, its contribution could be plotted as a function of $\Delta$ analogously to Figure \ref{plot-tensor}. This is not the case for long multiplets with nilpotent superconformal invariants. In these superconformal blocks, the coefficients of bosonic blocks that appear are theory dependent \cite{klps14, ls16, cls17}. Even before these exercises are done, it is clear that (\ref{sum-rule}) will not allow us to see individual cancellation within a given block. As stated earlier, our current form of the sum rule is contaminated by counterterms such that even the identity --- the supermultiplet with no descendants at all --- contributes a finite piece. This should be addressed in any serious attempt to study the structure of the OPE applicable to the beta function.

\section{Evolution equations}
To begin exploring the landscape of conformal manifolds, the sum rule (\ref{sum-rule}) is the most natural starting point. We would like to emphasize, however, that CFTs with exactly marginal deformations obey a much larger set of constraints. If a continuous line of theories has a solution at one point, these constraints are in principle enough to solve for local operators in the theories at all other points. The idea is to flow along the manifold in a given direction by applying the first order shifts
\begin{eqnarray}
\delta \Delta_i &=& -\delta g S_{d - 1} \lambda_{ii\hat{\mathcal{O}}} \nonumber \\
\delta \lambda_{ijk} &=& \delta g \biggl. \int_{\mathcal{R}} \textup{d}^dx \left < \mathcal{O}_i(0) \hat{\mathcal{O}}(x) \mathcal{O}_j(\hat{e}) \mathcal{O}_k(\infty) \right > \biggl |_{\mathrm{reg}} + \; \mathrm{perms} \label{ode-finite}
\end{eqnarray}
repeatedly.\footnote{Strictly speaking, the OPE coefficient $\lambda_{ii\hat{\mathcal{O}}}$ should be summed over all tensor structures if $\mathcal{O}_i$ is not a scalar.} The permutations denote integrals over $\mathcal{R}$ with different $(i, j, k)$ orderings. As discussed in the last section, this is equivalent to integrating the first permutation over all of space. Because the marginal operator takes us from one CFT to another, we will always be able to use the conformal block expansion on the right hand side of (\ref{ode-finite}). This allows us to write the results of conformal perturbation theory in exponentiated form:
\begin{eqnarray}
\frac{\textup{d}\Delta_i}{\textup{d}g} &=& -S_{d - 1} \lambda_{ii\hat{\mathcal{O}}} \nonumber \\
\frac{\textup{d}\lambda_{ijk}}{\textup{d}g} &=& \sum_{\mathcal{O}} \lambda_{i\hat{\mathcal{O}}\mathcal{O}} \lambda_{jk\mathcal{O}} \biggl. \int_{\mathcal{R}} \textup{d}^dx \frac{G_{\mathcal{O}}(x)}{|x|^{\Delta_i + d}} \biggl |_{\mathrm{reg}} + \; \mathrm{perms} \label{ode-infinitesimal} \; .
\end{eqnarray}
The sum rule from the last section is the special case found by taking $\mathcal{O}_i$, $\mathcal{O}_j$ and $\mathcal{O}_k$ to be $\hat{\mathcal{O}}$ itself.

Following OPE coefficients along the manifold is only meaningful if the deformation by $\hat{\mathcal{O}}$ preserves the starting normalization of all operators. In the absence of level crossing and other non-generic behaviour, it is possible to achieve this, but only in a particular scheme. This is, of course, the one we have been using which subtracts power-law divergences in the OPE. As stated in \cite{ks16}, other schemes --- which differ by a finite part --- can be recast into this language if we modify OPEs involving $\hat{\mathcal{O}}$ to include a contact term:
\begin{equation}
\hat{\mathcal{O}}(x) \mathcal{O}_i(0) \sim \alpha \delta(x) \mathcal{O}_i (0) \; . \label{contact1}
\end{equation}
It is clear from this that operator norms will shift by an $O(\alpha)$ amount and therefore be scheme dependent. Although we focus on non-degenerate spectra in this work, the situation becomes more interesting when we are allowed to write contact terms of the form:
\begin{equation}
\hat{\mathcal{O}}(x) \mathcal{O}_i(0) \sim \alpha^j_i \delta(x) \mathcal{O}_j (0) \; . \label{contact2}
\end{equation}
In this case, the field redefinitions needed to return to the initial normalization include not only rescalings but linear combinations within a degenerate subspace. While it is still possible to use a scheme that removes such mixing at a point, there is no guarantee that we may do so globally \cite{rsz94, bnp17}.

Clearly, the infinite coupled system (\ref{ode-infinitesimal}) will need to be truncated if one hopes to use it numerically. It is also unrealistic to expect a subset of the equations in (\ref{ode-infinitesimal}) to close among themselves. In particular, this means that one will have to compute the trajectories of OPE coefficients $\lambda_{ijk}(g)$ even if she is only interested in the dimensions $\Delta_i(g)$. It also means that CFT data involving only scalars will still receive contributions from spinning conformal blocks. For this reason, (\ref{ode-infinitesimal}) is most readily accessible in $d = 1$.

\subsection{One dimension}
\begin{figure}[h]
\centering
\includegraphics[scale=0.4]{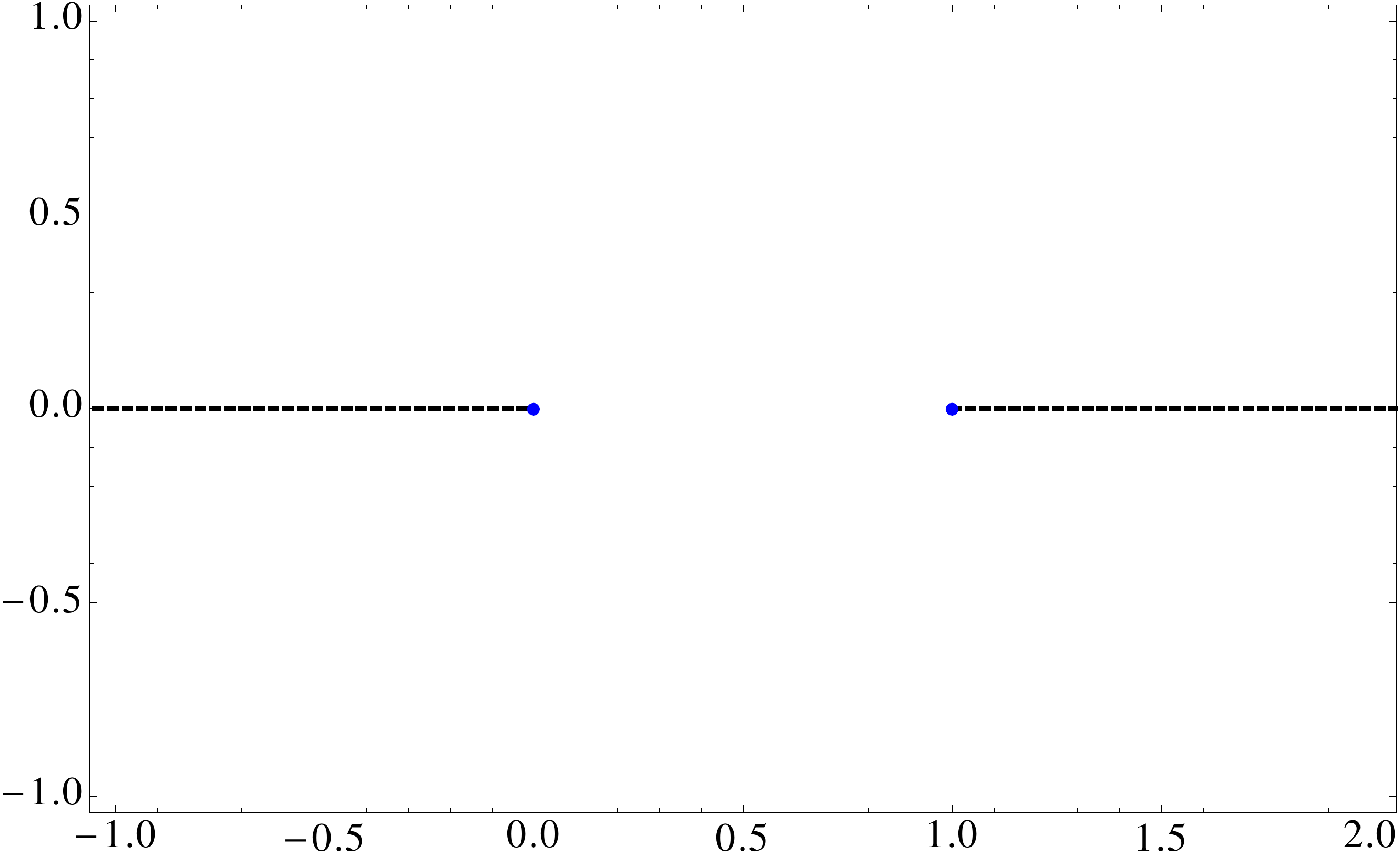}
\caption{Continuing to complex $z$, our blocks have one branch cut from $-\infty$ to $0$ and another from $1$ to $\infty$. This differs from higher-dimensional blocks which are analytic on $\mathbb{C} \setminus (1, \infty)$. As an example, we may multiply two $SL(2 ; \mathbb{R})$ blocks to get an $SL(2 ; \mathbb{C})$ block. This causes the left cut to cancel and the right cut to double.}
\end{figure}
Unlike in higher dimensions, there is only one cross-ratio on which 1D conformal blocks can depend. We take this to be $z = \frac{x_{12} x_{34}}{x_{13} x_{24}}$. The explicit functions
\begin{equation}
G_{\mathcal{O}}(z) = z^\Delta {}_2F_1(\Delta - \Delta_{12}, \Delta + \Delta_{34}, 2\Delta ; z) \; , \; x_1 < x_2 < x_3 < x_4 \label{block1}
\end{equation}
were found in \cite{do11, jp12}. Because the region of validity does not cover all of $\mathcal{R}$, we will need
\begin{equation}
\tilde{G}_{\mathcal{O}}(z) = \left ( \frac{z}{z - 1} \right )^\Delta {}_2F_1 \left ( \Delta  + \Delta_{12}, \Delta + \Delta_{34}, 2\Delta ; \frac{z}{z - 1} \right ) \; , \; x_2 < x_1 < x_3 < x_4 \label{block2}
\end{equation}
as well. Using (\ref{block1}) for $z \in (0, \frac{1}{2})$ and (\ref{block2}) for $z \in (-1, 0)$ leads to six regions. These correspond to the six possible orderings of $x_1, x_2, x_3$ after fixing $x_4 = \infty$.\footnote{This phenomenon of the s-channel splitting into two pieces occurs because there is no continuous way to move one operator around another. One consequence of this is that only cyclic permutations of $(i, j, k)$ leave $\lambda_{ijk}$ invariant \cite{m16}.} Putting back kinematic factors, we have
\begin{eqnarray}
\frac{\textup{d}\lambda_{ijk}}{\textup{d}g} &=& \sum_{\mathcal{O}} \left ( \lambda_{i\hat{\mathcal{O}}\mathcal{O}} \lambda_{jk\mathcal{O}} I_1 + \lambda_{\hat{\mathcal{O}}i\mathcal{O}} \lambda_{jk\mathcal{O}} I_2 \right ) + \mathrm{perms} \nonumber \\
I_1 &=& \int_0^{\frac{1}{2}} \textup{d}z \frac{G_{\mathcal{O}}(z)}{z^{\Delta_i + 1}} \nonumber \\
I_2 &=& \int_{-1}^0 \textup{d}z \frac{\tilde{G}_{\mathcal{O}}(z) (1 - z)^{\Delta_k - \Delta_j}}{(-z)^{\Delta_i + 1}} \; . \label{two-integrals}
\end{eqnarray}
A straightforward calculation yields
\begin{eqnarray}
I_1 &=& \biggl. \frac{z^{\Delta - \Delta_i}}{\Delta - \Delta_i} {}_2F_1(\Delta - \Delta_i, \Delta + \Delta_j - \Delta_k, 2\Delta, z) \biggl |_0^{\frac{1}{2}} \nonumber \\
&\rightarrow& \frac{2^{\Delta_i - \Delta}}{\Delta - \Delta_i} {}_2F_1 \left ( \Delta - \Delta_i, \Delta + \Delta_j - \Delta_k, 2\Delta, \frac{1}{2} \right ) \label{integral1}
\end{eqnarray}
where we have dropped finitely many terms in the last step. This amounts to renormalization as the negative powers of $z$ are precisely the divergences from $\Delta < \Delta_i$ operators in $\mathcal{O}_i \times \hat{\mathcal{O}}$. These are implicitly assumed to be subtracted in the naive integrals (\ref{two-integrals}).

To proceed further, we will need some identities, namely the two Pfaff transformations which combine to give an Euler transformation.
\begin{eqnarray}
{}_2F_1(a, b, c, z) &=& (1 - z)^{-a} {}_2F_1 \left ( a, c - b, c ; \frac{z}{z - 1} \right ) \nonumber \\
&=& (1 - z)^{-b} {}_2F_1 \left ( c - a, b, c ; \frac{z}{z - 1} \right ) \nonumber \\
&=& (1 - z)^{c - a - b} {}_2F_1(c - a, c - b, c ; z) \label{identities}
\end{eqnarray}
Using these on the second integral,
\begin{equation}
I_2 = \int_{-1}^0 \textup{d}z (-z)^{\Delta - \Delta_i - 1} {}_2F_1 ( \Delta - \Delta_i + 1, \Delta + \Delta_j - \Delta_k, 2\Delta ; z ) \; . \label{pre-integral2}
\end{equation}
We now renormalize and use a Pfaff transformation to send the argument $-1$ back to $\frac{1}{2}$.
\begin{equation}
I_2 \rightarrow \frac{2^{\Delta_i - \Delta}}{\Delta - \Delta_i} {}_2F_1 \left ( \Delta - \Delta_i, \Delta - \Delta_j + \Delta_k, 2\Delta ; \frac{1}{2} \right ) \label{integral2}
\end{equation}
We might have expected this form because the switch $(x_1, \Delta_1) \leftrightarrow (x_2, \Delta_2)$ is the same as $(x_3, \Delta_3) \leftrightarrow (x_4, \Delta_4)$. We arrive at our full coupled system by permuting labels in the results above.

\hspace*{-1.8cm}\vbox{\begin{empheq}[box=\widefbox]{align}
\frac{\textup{d} \Delta_i}{\textup{d}g} &= -2 \lambda_{ii\hat{\mathcal{O}}} \label{odes} \\
\frac{\textup{d} \lambda_{ijk}}{\textup{d}g} &= \sum_{\mathcal{O}} \frac{2^{\Delta_i - \Delta}}{\Delta - \Delta_i} \left [ \lambda_{i\hat{\mathcal{O}}\mathcal{O}} \lambda_{jk\mathcal{O}} {}_2F_1 \left ( \begin{tabular}{c} $\Delta - \Delta_i, \Delta + \Delta_{jk}$ \\ $2\Delta$ \end{tabular} ; \frac{1}{2} \right ) + \lambda_{\hat{\mathcal{O}}i\mathcal{O}} \lambda_{jk\mathcal{O}} {}_2F_1 \left ( \begin{tabular}{c} $\Delta - \Delta_i, \Delta + \Delta_{kj}$ \\ $2\Delta$ \end{tabular} ; \frac{1}{2} \right ) \right ] \nonumber \\
&+ \sum_{\mathcal{O}} \frac{2^{\Delta_j - \Delta}}{\Delta - \Delta_j} \left [ \lambda_{j\hat{\mathcal{O}}\mathcal{O}} \lambda_{ik\mathcal{O}} {}_2F_1 \left ( \begin{tabular}{c} $\Delta - \Delta_j, \Delta + \Delta_{ik}$ \\ $2\Delta$ \end{tabular} ; \frac{1}{2} \right ) + \lambda_{\hat{\mathcal{O}}j\mathcal{O}} \lambda_{ik\mathcal{O}} {}_2F_1 \left ( \begin{tabular}{c} $\Delta - \Delta_j, \Delta + \Delta_{ki}$ \\ $2\Delta$ \end{tabular} ; \frac{1}{2} \right ) \right ] \nonumber \\
&+ \sum_{\mathcal{O}} \frac{2^{\Delta_k - \Delta}}{\Delta - \Delta_k} \left [ \lambda_{k\hat{\mathcal{O}}\mathcal{O}} \lambda_{ij\mathcal{O}} {}_2F_1 \left ( \begin{tabular}{c} $\Delta - \Delta_k, \Delta + \Delta_{ij}$ \\ $2\Delta$ \end{tabular} ; \frac{1}{2} \right ) + \lambda_{\hat{\mathcal{O}}k\mathcal{O}} \lambda_{ij\mathcal{O}} {}_2F_1 \left ( \begin{tabular}{c} $\Delta - \Delta_k, \Delta + \Delta_{ji}$ \\ $2\Delta$ \end{tabular} ; \frac{1}{2} \right ) \right ] \nonumber
\end{empheq}}

\vspace{-4cm}
When $\Delta = \Delta_i$ is exchanged, it is helpful to use the formula
\begin{equation}
\biggl. \frac{\partial}{\partial \gamma} 2^{-\gamma} {}_2F_1 \left ( \begin{tabular}{c} $\gamma, \Delta + \Delta_{jk}$ \\ $2\Delta$ \end{tabular} ; \frac{1}{2} \right ) \biggl |_{\gamma = 0} = -\log(2) + \frac{\Delta + \Delta_{jk}}{4\Delta} {}_3F_2 \left ( \begin{tabular}{c} $1, 1, \Delta + \Delta_{jk} + 1$ \\ $2, 2\Delta + 1$ \end{tabular} ; \frac{1}{2} \right ) \; . \label{more-renormalized}
\end{equation}
There are two pertinent comments to be made about these equations. The first is that, as with the sum rule, we have made a choice for how to remove power-law divergences associated with individual conformal blocks. Our previous discussion, regarding Figure \ref{circle1} and Figure \ref{circle2}, is therefore applicable to the ODE system as well. We could, in principle, find other expressions to replace (\ref{odes}) that are just as correct. The predictions of these equations would only differ from those of (\ref{odes}) if they were used to evolve an initial condition that did not correspond to a CFT with an exactly marginal coupling.

The second is that we have not yet assumed invariance under parity. As there is no other dimension, we could just as well call it time-reversal. In a 1D parity-violating theory, it is often said that the dynamical part of the four-point correlator is not only a function of the cross-ratio $z$ \cite{b17}. This means that even for identical operators, $\left < \mathcal{O}(x_1)\mathcal{O}(x_2)\mathcal{O}(x_3)\mathcal{O}(x_4) \right >$ with the points sent to $(0, z, 1, \infty)$ need not have any kinematical relation to the same correlator with the points sent to $(z, 0, 1, \infty)$. In our derivation above, when we wrote each integrand in terms of the position of $\hat{\mathcal{O}}$, we did not assume that these could be glued together in a parity-preserving way. Doing so would amount to a restriction on the OPE coefficients. Consider a partial wave in the expansion of $\left < \mathcal{O}_1(x_1)\mathcal{O}_2(x_2)\mathcal{O}_3(x_3)\mathcal{O}_4(x_4) \right >$:
\begin{equation}
\lambda_{12\mathcal{O}}\lambda_{34\mathcal{O}} z^{\Delta - \Delta_1 - \Delta_2} {}_2F_1(\Delta - \Delta_1 + \Delta_2, \Delta + \Delta_3 - \Delta_4; 2\Delta; z) \; . \label{wave-before-parity}
\end{equation}
Under $(12)(34)$, $z$ maps to itself but (\ref{wave-before-parity}) maps to
\begin{equation}
\lambda_{21\mathcal{O}}\lambda_{43\mathcal{O}} z^{\Delta - \Delta_1 - \Delta_2} (1 - z)^{\Delta_1 - \Delta_2 - \Delta_3 + \Delta_4} {}_2F_1(\Delta + \Delta_1 - \Delta_2, \Delta - \Delta_3 + \Delta_4; 2\Delta; z) \; . \label{wave-after-parity}
\end{equation}
which suggests the Euler transformation. We see that (\ref{wave-after-parity}) is only the same function if $\lambda_{ijk}$ and $\lambda_{jik}$ differ at most by a sign. More generally, parity-preserving four-point functions must be invariant under $\{(), (12)(34), (13)(24), (14)(23)\} = \mathbb{Z}_2 \times \mathbb{Z}_2$ --- the set of transformations that stabilize $z$. This reflects the fact that there are 6 channels, but 24 configurations of four points. A non-trivial check is that, when OPE coefficients are restricted as above, our differential equation for $\lambda_{ijk}$ has the $S_4 / (\mathbb{Z}_2 \times \mathbb{Z}_2) = S_3$ symmetry befitting three operators that can couple to the parity-even $\hat{\mathcal{O}}$. Parity symmetry is common in defect CFTs, for example, because it is inherited from the parent theory.

\subsection{Avoided level crossing}
Even if one does not have a concrete model of a non-trivial conformal manifold in one dimension, a general prediction that can be distilled from (\ref{odes}) is a strong preference against nearly degenerate operators. Consider two primaries $\mathcal{O}_1$ and $\mathcal{O}_2$ where, for some value of the coupling $g_0$, $\Delta_1 - \Delta_2$ is parametrically small. We take it to be positive without loss of generality. By the von Neumann-Wigner non-crossing rule, we expect it to stay positive for $g > g_0$. For a demonstration of this occuring in maximally supersymmetric Yang-Mills theory, see \cite{brv13}. Also, \cite{k16} discussed this in the context of the $\frac{1}{N}$ expansion where the ``distance of closest approach'' is determined by the mixing matrix between planar and non-planar eigenstates. Taking the second derivative of one of the dimensions,
\begin{eqnarray}
\frac{\textup{d}^2 \Delta_1}{\textup{d}g^2} &=& -2 \frac{\textup{d}\lambda_{11\hat{\mathcal{O}}}}{\textup{d}g} \nonumber \\
&=& -4 \sum_{\mathcal{O}} \frac{2^{\Delta_1 - \Delta}}{\Delta - \Delta_1} \lambda^2_{1\hat{\mathcal{O}}\mathcal{O}} {}_2F_1 \left ( \begin{tabular}{c} $\Delta - \Delta_1, \Delta + \Delta_1 - 1$ \\ $2\Delta$ \end{tabular} ; \frac{1}{2} \right ) \nonumber \\
&-4& \sum_{\mathcal{O}} \frac{2^{\Delta_1 - \Delta}}{\Delta - \Delta_1} \lambda_{\hat{\mathcal{O}}1\mathcal{O}} \lambda_{1\hat{\mathcal{O}}\mathcal{O}} {}_2F_1 \left ( \begin{tabular}{c} $\Delta - \Delta_1, \Delta - \Delta_1 + 1$ \\ $2\Delta$ \end{tabular} ; \frac{1}{2} \right ) \nonumber \\
&-4& \sum_{\mathcal{O}} \frac{2^{1 - \Delta}}{\Delta - 1} \lambda_{\hat{\mathcal{O}}\hat{\mathcal{O}}\mathcal{O}} \lambda_{11\mathcal{O}} {}_2F_1 \left ( \begin{tabular}{c} $\Delta - 1, \Delta$ \\ $2\Delta$ \end{tabular} ; \frac{1}{2} \right ) \; . \label{level-start}
\end{eqnarray}
Our ability to quickly take the second derivative is only guaranteed because we are considering a manifold that has only one marginal operator and therefore a flat connection \cite{rsz94, bnp17}. Three values of $\Delta$ lead to small denominators in (\ref{level-start}) but the only one that can dominate the sum is $\Delta = \Delta_2$. Indeed, $\Delta_1$ and $1$ yield divergences whose renormalized versions are $O(1)$ numbers in view of (\ref{more-renormalized}). Knowing this, we can approximate the differential equation as
\begin{eqnarray}
\frac{\textup{d}^2 \Delta_1}{\textup{d}g^2} &\approx& 4 \frac{\lambda^2_{12\hat{\mathcal{O}}} 2^{\Delta_1 - \Delta_2}}{\Delta_1 - \Delta_2} \left [ {}_2F_1 \left ( \begin{tabular}{c} $-\Delta_{12}, \Delta_2 + \Delta_1 - 1$ \\ $2\Delta_2$ \end{tabular} ; \frac{1}{2} \right ) + {}_2F_1 \left ( \begin{tabular}{c} $-\Delta_{12}, \Delta_2 - \Delta_1 + 1$ \\ $2\Delta_2$ \end{tabular} ; \frac{1}{2} \right ) \right ] \nonumber \\
&\approx& \frac{8 \lambda^2_{12\hat{\mathcal{O}}}}{\Delta_1 - \Delta_2} \; . \label{level-one}
\end{eqnarray}
In the first step, we have assumed that our theory has a parity symmetry. Since operators with different quantum numbers are allowed to cross, we are interested in $\mathcal{O}_1$ and $\mathcal{O}_2$ which have the same parity. This means that $\lambda_{11\hat{\mathcal{O}}}$, $\lambda_{22\hat{\mathcal{O}}}$ and $\lambda_{12\hat{\mathcal{O}}}$ all exist and are independent of the label ordering. After switching the labels to find $\frac{\textup{d}^2 \Delta_2}{\textup{d}g^2}$, the differential equation for $y \equiv \Delta_1 - \Delta_2$ is
\begin{equation}
\frac{\textup{d}^2y}{\textup{d}g^2} = \frac{16\lambda^2_{12\hat{\mathcal{O}}}}{y} \label{level-y1} \; .
\end{equation}
It is not necessarily true that $\lambda_{12\hat{\mathcal{O}}}$ is slowly varrying compared to $y$. Therefore we will compute its variation by going back to (\ref{odes}).
\begin{eqnarray}
\frac{\textup{d}\lambda_{12\hat{\mathcal{O}}}}{\textup{d}g} &\approx& \frac{2}{\Delta_2 - \Delta_1} \lambda_{12\hat{\mathcal{O}}} \lambda_{22\hat{\mathcal{O}}} + \frac{2}{\Delta_1 - \Delta_2} \lambda_{12\hat{\mathcal{O}}} \lambda_{11\hat{\mathcal{O}}} \nonumber \\
&=& \frac{2 \lambda_{12\hat{\mathcal{O}}}}{y} (\lambda_{11\hat{\mathcal{O}}} - \lambda_{22\hat{\mathcal{O}}}) \nonumber \\
&=& -\frac{\lambda_{12\hat{\mathcal{O}}}}{y} \frac{\textup{d}y}{\textup{d}g} \; . \label{separable}
\end{eqnarray}
Solving this separable equation yields
\begin{equation}
\frac{\textup{d}^2y}{\textup{d}g^2} = \frac{2c}{y^3} \label{level-y2}
\end{equation}
where $c$ is a positive constant. The right hand side is $-\frac{\textup{d}}{\textup{d}y} U(y)$ for the potential function $U(y) = \frac{c}{y^2}$. Having turned this into a 1D classical scattering problem, we see that $U(y)$ is always a repulsive potential. In particular, $U(0) = +\infty$ means that a shrinking $y$ will always reach a turning point.

This argument appears not to rely on one dimension. The crucial feature we have used is that $\frac{\textup{d}\lambda_{ijk}}{\textup{d}g}$ contains a $\frac{1}{\Delta - \Delta_i}$ singularity for each dimension $\Delta$ primary in $\mathcal{O}_i \times \hat{\mathcal{O}}$. To see this appearing more generally, consider the regulated integral
\begin{equation}
\int_{B(r)} \textup{d}^dx \left < \mathcal{O}_i(0) \hat{\mathcal{O}}(x) \mathcal{O}_j(\hat{e}) \mathcal{O}_k(\infty) \right > \label{integral-before-insertion}
\end{equation}
where $r \ll 1$. Even if we cannot evaluate (\ref{integral-before-insertion}) exactly, we know that one term of it will come from inserting the pair of states $\left | \mathcal{O} \right > \left < \mathcal{O} \right |$. It can then be argued that operators with $\Delta \rightarrow \Delta_i$ will dominate over all other choices. In the all-scalar case for instance, the $\mathcal{O}$ term of the integral becomes
\begin{equation}
\int_{B(r)} \textup{d}^dx \frac{\lambda_{i\hat{\mathcal{O}}\mathcal{O}}\lambda_{jk\mathcal{O}}}{|x|^{d + \Delta_i - \Delta}} \label{integral-after-insertion}
\end{equation}
which leads to $\frac{r^{\Delta - \Delta_i}}{\Delta - \Delta_i}$. Depending on the sign of $\Delta - \Delta_i$, the boundary term is either manifestly zero, or zero as a result of our regulating procedure.

Besides demonstrating agreement with \cite{k16}, lack of level crossing may be seen as a consistency check of the equations we have been using. If operator pairs with $\Delta_1 \approx \Delta_2$ could occur throughout the conformal manifold, the differential equation for $\lambda_{11\hat{\mathcal{O}}}$ would contain terms that are large --- possibly large enough to combat the suppression by powers of $g$ in conformal perturbation theory. From this point of view, it is encouraging that our system abhors the regime $y \ll 1$.

\subsection{The compact free boson}
Despite our success in the previous calculation, the question of whether evolution equations can be used to solve a conformal manifold is still open. We are therefore asking whether (\ref{ode-infinitesimal}) contains a full description of the operator algebra, including non-perturbative dynamics. As a first indication that this is plausible, we will show that the compact free boson satisfies the desired ODE system exactly.

If the free boson theory is written in terms of a fundamental field $\phi(z, \bar{z})$, we may use the equation of motion to split it into holomorphic and anti-holomorphic parts; $\phi(z, \bar{z}) = X(z) + \bar{X}(\bar{z})$. When the fields take values on a circle of radius $r$, this leads to the on-shell action
\begin{equation}
S = -\frac{1}{\pi} \int \textup{d}z \textup{d}\bar{z} \; \partial X \bar{\partial} \bar{X} \;\;\; , \;\;\; X \sim X + 2\pi r \; . \label{free-action}
\end{equation}
The prefactor is chosen so that the $U(1)$ current, $J(z) \equiv i\partial X(z)$, is unit normalized. We have already seen that the marginal operator in this model is $\hat{\mathcal{O}} = J\bar{J}$. Since this is proportional to the Lagrangian density,
\begin{equation}
S + \delta g \int \textup{d}z \textup{d}\bar{z} \; \hat{\mathcal{O}} = (1 + \pi \delta g) S \; . \label{shifted-action}
\end{equation}
To get back to the original action, we must rescale $X$ and $\bar{X}$ by $\sqrt{1 + \pi \delta g}$. This shifts the compactification radius by $\delta r = \frac{1}{2} \pi r \delta g$, allowing us to solve
\begin{equation}
r = r_0 e^{\frac{\pi}{2} g} \; . \label{radius-action}
\end{equation}

It is enough to check that (\ref{ode-infinitesimal}) is satisfied for the $U(1)$ primaries. These are the $V_{q, \bar{q}}(z, \bar{z}) = :e^{i\sqrt{2}qX(z)} e^{i\sqrt{2}\bar{q}\bar{X}(\bar{z})}:$ vertex operators. By periodicity and single-valuedness, the charges are specified by two integers \cite{bs93}
\begin{equation}
(\sqrt{2} q, \sqrt{2} \bar{q}) \in \Gamma = \left \{ \left ( \frac{n}{r} + \frac{mr}{2}, \frac{n}{r} - \frac{mr}{2} \right ) \; \middle | \; m, n \in \mathbb{Z} \right \} \; . \label{charge-lattice}
\end{equation}
Combining (\ref{radius-action}) with (\ref{charge-lattice}), we find
\begin{eqnarray}
\frac{\textup{d}\Delta_{(m, n)}}{\textup{d}g} &=& \frac{\textup{d}}{\textup{d}g} \left ( q^2 + \bar{q}^2 \right ) \nonumber \\
&=& -\pi \left ( \frac{n^2}{r^2} - \frac{m^2 r^2}{4} \right ) \; . \label{eq1-lhs}
\end{eqnarray}
To see if our equations predict this, we must evaluate $\left < \hat{\mathcal{O}}(z, \bar{z})  V_{(m, n)}(z_1, \bar{z}_1) V^*_{(m, n)}(z_2, \bar{z}_2) \right >$. This is simple because the Virasoro primary $\hat{\mathcal{O}}$ is in the identity multiplet of $U(1)$. Using the fact that $J = a_{-1} I$, the Ward identities
\begin{eqnarray}
J(z) V_{q, \bar{q}}(w, \bar{w}) = \frac{q}{z - w} V_{q, \bar{q}}(w, \bar{w}) + \dots \nonumber \\
\bar{J}(\bar{z}) V_{q, \bar{q}}(w, \bar{w}) = \frac{\bar{q}}{\bar{z} - \bar{w}} V_{q, \bar{q}}(w, \bar{w}) + \dots \label{free-ward}
\end{eqnarray}
are easy to derive. This leads to the OPE coefficient
\begin{eqnarray}
\lambda_{(m, n)(m, n)\hat{\mathcal{O}}} &=& q\bar{q} \nonumber \\
&=& \frac{1}{2} \left ( \frac{n^2}{r^2} - \frac{m^2 r^2}{4} \right ) \; . \label{eq1-rhs}
\end{eqnarray}
The first equation in (\ref{ode-infinitesimal}) is therefore verified.

Turning to the second equation, the left hand side vanishes by the well known OPE of vertex operators:
\begin{equation}
\lambda_{(m_1, n_1)(m_2, n_2)(m_3, n_3)} = \delta_{q_1 + q_2 + q_3, 0} \delta_{\bar{q_1} + \bar{q_2} + \bar{q_3}, 0} \; . \label{eq2-lhs}
\end{equation}
The right hand side, while daunting for a general 2D theory, may be evaluated for the free boson without the conformal block expansion. This is possible because of the Ward identity and the formula
\begin{equation}
\biggl. \int \textup{d}^2z \frac{1}{z - z_i} \frac{1}{\bar{z} - \bar{z}_j} \biggl |_{\mathrm{reg}} =
\begin{cases}
2\pi \log |z_{ij}| & i \neq j \\
0 & i = j
\end{cases} \label{log-pv}
\end{equation}
understood as a principal value.\footnote{One way to see that (\ref{log-pv}) holds is to check that both sides are Green's functions of the operators $\partial_i \bar{\partial}_i$ and $\partial_j \bar{\partial}_j$.} We now follow the method of \cite{fr03} to show that a first order insertion of $\hat{\mathcal{O}}$ produces no shift in the highest-weight OPE coefficients.
\begin{eqnarray}
&& \biggl. \int \textup{d}^2z \left < J(z) \bar{J}(\bar{z}) V_{q_1, \bar{q}_1}(z_1, \bar{z}_1) V_{q_2, \bar{q}_2}(z_2, \bar{z}_2) V_{q_3, \bar{q}_3}(z_3, \bar{z}_3) \right > \biggl |_{\mathrm{reg}} \nonumber \\
&=& \biggl. \int \textup{d}^2z \left ( \frac{q_1}{z - z_1} + \frac{q_2}{z - z_2} + \frac{q_3}{z - z_3} \right ) \left ( \frac{\bar{q}_1}{\bar{z} - \bar{z}_1} + \frac{\bar{q}_2}{\bar{z} - \bar{z}_2} + \frac{\bar{q}_3}{\bar{z} - \bar{z}_3} \right ) \biggl |_{\mathrm{reg}} \label{eq2-rhs} \\
&& \left < V_{q_1, \bar{q}_1}(z_1, \bar{z}_1) V_{q_2, \bar{q}_2}(z_2, \bar{z}_2) V_{q_3, \bar{q}_3}(z_3, \bar{z}_3) \right > \nonumber \\
&=& 2\pi [(q_1\bar{q}_2 + q_2\bar{q}_1) \log |z_{12}| - (2q_1\bar{q}_1 + q_1\bar{q}_2 + q_2\bar{q}_1) \log |z_{13}| - (2q_2\bar{q}_2 + q_1\bar{q}_2 + q_2\bar{q}_1) \log |z_{23}|] \nonumber \\
&& \left < V_{q_1, \bar{q}_1}(z_1, \bar{z}_1) V_{q_2, \bar{q}_2}(z_2, \bar{z}_2) V_{q_3, \bar{q}_3}(z_3, \bar{z}_3) \right > \nonumber
\end{eqnarray}
In the last step, we have used the fact that $q_3$ is a shorthand for $-(q_1 + q_2)$ and similarly for $\bar{q}_3$. This prefactor, with its $\log |z_{ij}|$ terms, is precisely what we would find if we applied $\frac{\textup{d}}{\textup{d}g}$ to the three-point function and allowed it to act on the charge lattice alone.

The derivation in \cite{fr03} applies not only to the free boson, but to all WZW models deformed by current-current terms. When the currents are non-abelian, the terms that generate the Cartan subalgebra stay exactly marginal. In other words, the conformal manifold is multi-dimensional. As this is an interesting topic for the future, it is likely that the work done on WZW models can serve as a guide for finding the appropriate generalization of (\ref{ode-infinitesimal}).

\section{Conclusion}
We expect our main results, (\ref{sum-rule}) and (\ref{odes}), to fit into a larger effort to understand theories with $\beta(g) = 0$ systematically. First, we intend to take a closer look at the question \textit{what is the contribution of a given primary operator to the two-loop beta function of a deformed CFT?} Our ability to discuss the sum rule in the context of known conformal manifolds was limited because this question does not have a unique answer. This arose from the lack of a clear prescription for distributing counterterms in the conformal block expansion. Given the recent interest in using conformally covariant basis functions other than the traditional blocks, it is possible that results in \cite{ss16, ps17, l17, gkss17a, gkss17b} could help ensure that we are using the right language.

Moving beyond the sum rule, we have seen a rich set of relations for how the local CFT data depend on the coupling. The toy model we have chosen avoids the mixing and interesting geometry induced by multiple couplings. We further restricted to $d = 1$ as it allowed us to write (\ref{odes}) in a form that had no unevaluated integrals. The demonstration of avoided level crossing was appealingly simple and we believe that extending this result to higher dimensions is purely a book-keeping exercise. A more involved goal is applying this framework to follow the full set of $\{ \Delta_i(g), \lambda_{ijk}(g) \}$ data in a realistic model numerically. Doing this for $d \geq 3$ would be difficult at present since there is still a gap in the literature concerning superconformal blocks with external spin. The $d = 2$ case seems tractable, but only if it involves Virasoro blocks. Treating $d = 2$ with global blocks would require us to diagonalize the two-point functions of degenerate operators before every $g \mapsto g + \delta g$ step. It is therefore of interest to find efficient algorithms for conformal blocks that lack a closed form, as explored in \cite{s11, lsswy15, mp17, p17}.

\section*{Acknowledgements}
I am very grateful to Leonardo Rastelli for providing the initial idea for this project and emphasizing the importance of understanding conformal manifolds from the bootstrap perspective. I thank him and Christopher Beem for offering much insight during the editing of this draft. Edoardo Lauria, Dalimil Maz\'{a}\v{c}, Marco Meineri, Carlo Meneghelli, Sabrina Pasterski, Miguel Paulos, Daniel Robbins, Slava Rychkov and Yifan Wang provided helpful discussions. Much of this work was completed at the Bootstrap 2017 workshop, generously hosted by the ICTP-SAIFR in S\~{a}o Paulo, Brazil.

\bibliographystyle{unsrt}
\bibliography{references}

\end{document}